\documentclass[prl,twocolumn,showpacs]{revtex4}      
\input epsf
\newcommand{\be}{\begin{equation}}
\newcommand{\ee}{\end{equation}}

\font\mybb=msbm10 at 11pt

\def\bb#1{\hbox{\mybb#1}}

\def\Z {\bb{Z}}

\def\ben{\begin{equation}}
\def\een{\end{equation}}
\def\bea{\begin{eqnarray}}
\def\eea{\end{eqnarray}}

\begin{document}

\title{Skyrmions with massive pions }
\author{Richard A. Battye$^{1}$ and Paul M. Sutcliffe$^{2}$}

\address{$^{1}$ Jodrell Bank Observatory, University of Manchester, 
Macclesfield, Cheshire SK11 9DL, U.K. \\ $^{2}$ Institute of Mathematics,
 University of Kent at Canterbury, Canterbury CT2 7NF, U.K.}


\begin{abstract}
In the Skyrme model with massless pions, the minimal energy multi-Skyrmions
are shell-like, with the baryon density localized on the
edges of a polyhedron that is approximately spherical and generically of the 
fullerene-type. 
In this paper we show that in the Skyrme model with
massive pions these configurations are
unstable for sufficiently large baryon number. Using numerical simulations
 of the full nonlinear field
theory, we show that these structures collapse to
form qualitatively different stable Skyrmion solutions. These new
Skyrmions have a flat structure and display a clustering
phenomenon into lower charge components, particularly components of
baryon numbers three and four. These new qualitative features of 
Skyrmions with massive pions are encouraging in comparison with the
expectations based on real nuclei.   
\end{abstract}


\maketitle

The Skyrme model \cite{Sk} 
 is a nonlinear theory of pions which is
an approximate, low energy effective theory of quantum chromodynamics,
obtained in the limit of a large number of quark colours \cite{Wi}.
Skyrmions are topological soliton solutions of the model and 
are candidates for an effective description of nuclei, with an identification
between soliton and baryon numbers (for a review see \cite{book}).

Recent investigations \cite{BS10} have suggested that there will be important
qualitative differences between Skyrmions with massive and massless pions.
Minimal energy multi-Skyrmions with massless pions
are shell-like, with the baryon density localized on the
edges of a polyhedron that is approximately spherical and generically of the 
fullerene-type. However, for massive pions such solutions fail to be
bound states for particular baryon numbers, with a strong dependence
upon the value of the pion mass \cite{BS10}. It was suggested that there
must be qualitative changes as the pion mass is increased, but the 
form of the new Skyrmion solutions was not computed. In this paper we address
this issue by performing numerical simulations of the full nonlinear field
theory. Starting with the massless pion Skyrmions which we perturb in the 
massive theory, we find that the fullerene structures collapse to
form qualitatively different stable Skyrmions. The new
Skyrmions have a flat structure and display a clustering
phenomenon into lower charge components, particularly components of
baryon numbers three and four. These new qualitative features 
are encouraging in comparison with the expectations based on real nuclei,
and provide motivation for a number of further investigations.    

The field $U$ of the Skyrme model \cite{Sk} 
is an $SU(2)$-valued scalar with an associated  current
 $R_i=(\partial_i U)U^\dagger.$
In this paper we are only concerned with static fields, so we can
define the Skyrme model by its energy
\begin{eqnarray}
E=\frac{1}{12\pi^2}\int \large\{&-&{1 \over 2}\mbox{Tr}(R_iR_i)-{1 \over 16}
\mbox{Tr}([R_i,R_j][R_i,R_j])\cr &+&m^2\mbox{Tr}(1-U)\large\} \, d^3x\,,
\label{skyenergy}
\end{eqnarray}
which is normalized so that the Faddeev-Bogomolny bound reads $E\ge |B|,$
where $B$ is the baryon number (or topological charge) given by the 
degree of the mapping. In the above the energy and length units (which
must be fixed by comparison to real data) have been
scaled away, leaving only the pion mass parameter $m.$ This parameter
is proportional to the (tree-level) pion mass, in scaled units. 

The most detailed studies of multi-Skyrmions \cite{BS3} have assumed
massless pions ($m=0$), and only for very low baryon numbers have massive
pions been included \cite{BTC,BBT,Kop}. Furthermore, when a non-zero 
pion mass has been introduced, it has always been set to the experimentally
measured value determined by matching to the linearized pion theory.
This is the approach adopted in early studies of a single Skyrmion \cite{AN}
based on reproducing the masses of the proton and delta resonance
\cite{ANW}, and leads to the value $m=0.526.$ 
However, this approach has recently been re-examined \cite{BKS,HoMa1}
by removing some of the assumed approximations, with the result that
the proton and delta masses can only be reproduced if the pion mass is
taken to be larger than roughly twice the experimentally measured value. 
One interpretation of this development is that the pion mass parameter 
in the Skyrme model
should be regarded as a renormalized pion mass, and therefore not be
fixed by the experimentally measured value, but instead treated 
as a free parameter to be adjusted to best reproduce the properties of
nuclei. As we are mainly interested in qualitative phenomena in this paper,
and motivated by the results in Ref.\cite{BKS}, 
we shall set $m=1$ for most of our study. We shall briefy discuss how
our results are modified by alternative values, but an in-depth 
quantitative analysis of multi-Skyrmion properties
as a function of $m$ is beyond the scope of the current investigation.

For massless pions the minimal energy Skyrmions have been obtained
for all $B\le 22$ \cite{BS3}, and are well-approximated by the rational
map ansatz \cite{HMS}. This involves a decomposition of the field into
a radial and angular dependence, with the angular dependence determined
by a rational map between Riemann spheres. One of the reasons it is such
a good approximation is due to the roughly spherical shell-like distribution
of the energy density of these Skyrmions.

The vacuum value $U=1$ is attained at spatial infinity and the number 
of points in space (counted with multiplicity) at which the anti-vacuum
value $U=-1$ is attained must equal the baryon number $B,$ by simple
topological arguments. These anti-vacuum points are, therefore, a
useful characterization of the field, and in particular are the
locations of a set of single Skyrmions if they are all well-separated.
 In the rational map approximation all $B$ points are coincident
at the centre of the shell, and this appears to be a property shared by the
exact solutions for massless pions when there is a large amount of symmetry
(axial or Platonic, for example). However, when there is only dihedral symmetry
or less, these $B$ anti-vacuum points can split into $B$ distinct
points consistent with any dihedral symmetry. For example, the $B=9$
solution has $D_{4d}$ symmetry and one anti-vacuum point is at the origin
with the other eight on the vertices of a regular octagon in the plane
orthogonal to the main symmetry axis and containing the origin.
The locations of the anti-vacuum points in the case of massive pions
will be discussed below.

In order to test the stability of shell-like solutions 
we have used the same methods described in detail 
in Ref.\cite{BS3} to numerically relax field configurations to static
solutions which are local energy minima of the Skyrme model with $m=1.$ 
The results presented here used grids containing
$101^3$ points with a lattice spacing $dx=0.1,$ though other grid
sizes and lattice spacings were tested for comparison. 

As an initial condition we take the minimal energy charge $B$ Skyrmion
for massless pions (actually we use its rational map approximation, but
this is good enough) and perturb it by squashing it by $20\%$ in a direction
which is not aligned with any symmetry axis of the Skyrmion. This ensures
that the initial configuration has no exact symmetry. The results of
energy relaxations with $m=1$ on a parallel supercomputer are presented below.

For $B\le 9$ we find that essentially the same configurations as in
the massless case are recovered, though the Skyrmions are now smaller in
size and are exponentially (rather than algebraically) localized, which are
the obvious consequences of massive pions. There is spherical symmetry for
$B=1,$ axial symmetry for $B=2,$  Platonic symmetry for $B=3,4,7,$
and dihedral symmetry for $B=5,6,8,9.$ In the examples with dihedral symmetry
the Skyrmion is slightly squashed in the direction of the main symmetry
axis, so that it appears flat in comparison to the massless case. 
We shall elaborate on this a little later when we consider the $B=8$
example in more detail. For now we simply remark that the squashed nature
of the relaxed configuration is not related to the squashing used to perturb
the initial condition.

\begin{figure}[t]
\begin{center}
\leavevmode
\epsfxsize=7cm\epsffile{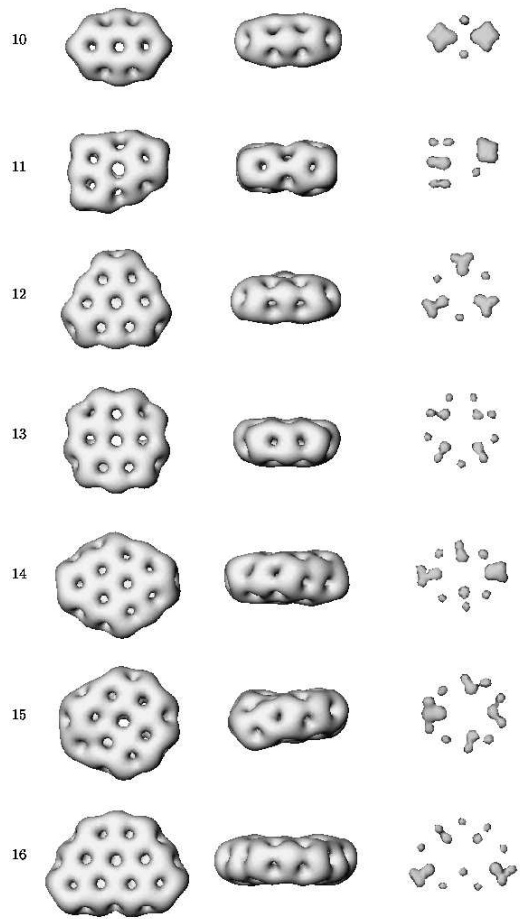}
\caption{Skyrmion solutions for $10\le B\le 16,$ with pion 
mass parameter $m=1.$ The first two columns are baryon density
isosurface plots from two different viewing angles, and the third column
displays isosurfaces where $\frac{1}{2}\mbox{Tr}(U)=-0.9.$ This allows
the identification of the anti-vacuum points and their cluster structure.
All plots are displayed using the same scale.}
\label{fig-10to16}
\vskip 0cm
\end{center}
\end{figure}

For $B\ge 10$ we find that the spherical shell-like configurations are unstable
and they relax to solutions which are much less symmetric and have a remarkably
flat structure. We present these solutions, to scale, in Fig.~\ref{fig-10to16}
for $10\le B\le 16.$ In this figure the first two columns are baryon density
isosurface plots from two different viewing angles, and the third column
displays isosurfaces where $\frac{1}{2}\mbox{Tr}(U)=-0.9.$ This allows us
to identify the anti-vacuum points, together with the nature of their 
clustering. Choosing an isosurface value closer to $-1$ 
allows us to pinpoint the anti-vacuum points more 
accurately (and in particular count them to confirm that there are $B$
in each case) but the information of how they group into clusters is then
lost, and this will be an important feature as we shall see below.   

The views from the top and the side, displayed in the first
two columns of Fig.~\ref{fig-10to16},
confirm the flat planar structure of these solutions.
The instability of spherical shells for massive pions originates from
the fact that the interior of the shell is a large volume over which
the Skyrme field is close to the anti-vacuum value $U=-1,$ and the
pion mass term makes this energetically unfavourable \cite{BS10}.
The new Skyrmion solutions reduce this energy contribution by 
being very flat and hence reducing the volume to surface area ratio
of the configuration.

Clearly, flat planar structures can have at most a dihedral symmetry,
so they will generally be less symmetric than the spherical structures,
which often have Platonic symmetries. The symmetries of the new Skyrmions 
are presented in Table~\ref{tab-sym}, where $\Z_2$ denotes only a
reflection symmetry. It can be seen from this Table that these Skyrmions
typically have very little symmetry, even for planar arrangements.

Also shown in  Table~\ref{tab-sym} is the energy per baryon $E/B$
for each solution. This demonstrates that the energy per baryon is
similar for all baryon numbers in this range, though there is a clear 
tendency for the energy to be slightly higher for odd baryon numbers.
For each of the odd baryon numbers in  Table~\ref{tab-sym} the
energy per baryon is higher than for the even baryon number solutions
obtained by either adding or subtracting one from the baryon number.
This is a new feature not shared by Skyrmions with massless pions,
where often it is solutions with odd baryon numbers that have 
unusually low energy \cite{BS3}.
This new feature is an encouraging sign in relating to real nuclei, where 
even mass numbers have larger binding energies and are more
abundant than those with odd mass numbers.

It should be noted that we have found it more difficult to accurately compute 
energies in the massive pion theory than in the massless one, 
since the pion mass produces a more rapid spatial variation of the fields. 
Thus some caution should be applied in trusting the third decimal place in 
the energies presented in Table~\ref{tab-sym}.
 
\begin{table}
\caption{\label{tab-sym}The symmetry group $G$ and
the energy per baryon $E/B,$ for Skyrmions 
with $10\le B\le 16$ and pion mass parameter $m=1.$}
\centering
\begin{ruledtabular}
\begin{tabular}{|c|c|c|c|c|c|c|c|}
$B$ &  10 & 11 & 12 & 13 & 14 & 15 &16 \\  \hline
$G$ & $D_{2h}$ & $\Z_2$ & $C_3$ & $\Z_2$ & $\Z_2$ & $C_2$ & $\Z_2$ \\ 
$E/B$ & 1.290  &  1.297 & 1.289  & 1.292 & 1.288 & 1.291 & 1.288 \\ 
\end{tabular}
\end{ruledtabular}
\end{table}


To interpret the structure of the solutions presented in 
Fig.~\ref{fig-10to16} it is helpful to consider the clustering of
the anti-vacuum points displayed in the third column.
For example, for $B=10$ there are two groups of four anti-vacuum points
and two single points. This suggests that the configuration should be
interpreted as being composed of two charge four Skyrmions and 
two single Skyrmions.
Recalling that the baryon density of the $B=4$ Skyrmion is localized 
on the edges of a cube, this interpretation is consistent with the
baryon density plot in the first column, where two deformed cube-like
structures are visible. 

As another example, the anti-vacuum points
of the $B=12$ Skyrmion are clustered in three groups of three, in
an equilateral triangle, with three single Skyrmions on the dual triangle.
This suggests that this $B=12$ Skyrmion contains three tetrahedra
(the minimal $B=3$ Skyrmion having its baryon density localized 
on the edges of a tetrahedron) and three single Skyrmions. This is
not so obvious from the baryon density isosurface presented in
the first column of Fig.~\ref{fig-10to16}, however, we were able
to confirm this interpretation as follows. A simulation of the full
nonlinear time dependent Skyrme equations was performed (using the 
numerical code
described in detail in Ref.\cite{BS3}) with an initial condition consisting
of the new $B=12$ Skyrmion, but with an overall size rescaling so that it
was initially too small. During the time evolution this configuration expands
and due to the large kinetic energy it breaks up into a triangular 
arrangement of three tetrahedra, plus the three single Skyrmions on
the dual triangle, in agreement with the prediction based on the anti-vacuum
points. Each tetrahedron has a face parallel to the plane of the
triangle and a vertex pointing up from this plane. This breaks the up-down
symmetry in the plane and explains why
the $B=12$ Skyrmion has only a $C_3$ symmetry, not a $D_3$ symmetry. 
From the baryon density figure in the first column of Fig.~\ref{fig-10to16}
it can be seen that there is a hole in the centre of the top of the Skyrmion,
but there is no corresponding hole in the centre at the bottom, as can
be confirmed by looking through the top hole.

The Skyrmions with other values of $B$  have a similar cluster structure,
with groups of three and four anti-vacuum points often occuring; suggesting
that substructures of charges three and four are favoured. This is another
encouraging development, since it is known that many nuclei may be
described as arrangements of alpha particles. This aspect will be
explored elsewhere \cite{BMS}.

 Given that 
these Skyrmions appear to be formed from combinations of smaller charge
units, it seems likely that there will be many local minima given by
different possible partitions and geometrical arrangements of the 
smaller charge components. 
Many of the Skyrmions shown in Fig.~\ref{fig-10to16} have very
little symmetry and this appears to be due to the fact that there is a
partial clustering together with several single anti-vacuum points.  
Thus it may be that these stable solutions are only local minima and
different global minima may exist which have more clustering. 
It is a computationally expensive task to investigate this issue, since
it requires many simulations for each baryon number using a variety
of different initial conditions. Such a study is currently underway
and the results will be reported in the near future \cite{BMS}.

\begin{figure}
\begin{center}
\leavevmode
\epsfxsize=8cm\epsffile{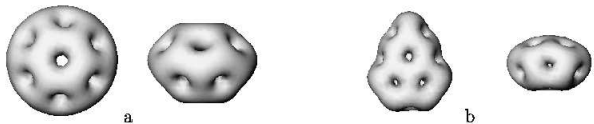}
\caption{Two views of the baryon density isosurface of the $B=8$ Skyrmion
(a) with $m=1$; (b) with $m=2.$}
\label{fig-8m1m2}
\vskip 0cm
\end{center}
\end{figure}

As discussed in detail in Ref.\cite{BS10}, as the pion mass is increased 
its effects become more important and hence the expectation is that 
shell-like Skyrmions (including squashed versions) will become unstable 
at lower baryon numbers. To test this conjecture we consider
the $B=8$ Skyrmion with pion mass parameter $m=1$ and $m=2.$
In Fig.~\ref{fig-8m1m2}(a) we present two views of the 
 baryon density isosurface of the $B=8$ Skyrmion with pion mass parameter
$m=1.$ It has $D_{6d}$ symmetry and, as mentioned
earlier, is essentially the same as in the massless case except for
the slight squashing in the direction of the main symmetry axis (which
can be seen in the second figure in Fig.~\ref{fig-8m1m2}(a)). 
In Fig.~\ref{fig-8m1m2}(b) we display two views of the baryon density
isosurface for the Skyrmion solution we find when $m=2.$ This is a flat
Skyrmion with only a $\Z_2$ reflection 
symmetry and is evidently of the same type as
the Skyrmions we have found for $B\ge 10$ with $m=1.$ The anti-vacuum points
divide into a group of five and another group of three. This confirms
 our expectation that a larger pion mass yields a qualitative change
in the structure and symmetry of Skyrmions at lower baryon numbers.

We have explorerd a range of different pion masses 
($m=\frac{1}{4},~\frac{1}{2},~2$)  in addition to the $m=1$ 
results presented in detail. 
Using the $m=1$ solutions as initial conditions, we found that
the solutions presented in Fig.~\ref{fig-10to16} are essentially
reproduced for $m=2$ and $m=\frac{1}{2},$   
but not for $m=\frac{1}{4}$ where the solutions 
reverted to those found for $m=0$.  

In summary, we have demonstrated that there are important qualitative 
differences between Skyrmions with massless and massive pions. This has
crucial implications for the ultimate long term aim of comparing the 
Skyrme model of nuclei with real data. The qualitative changes we have 
found suggest that there is an increased hope of reproducing some of
the most important properties of nuclei, and motivates further 
investigations of the properties of the solutions, some of which are currently
underway.   

In a zero mode quantization of Skyrmions the symmetry of the classical 
solution plays an important role in providing constraints on the 
spin, isospin and parity quantum numbers. For massless pions this has been 
studied in detail \cite{Kr2} for baryon numbers upto 22, with some 
success in comparing with experimental data, particularly for low baryon 
numbers. However, there are a number of discrepancies which arise
due to the large symmetry groups of the solutions.  
The results we have presented in this paper reveal that 
Skyrmions with massive pions are much less symmetric, so it would be
interesting to calculate the new constraints on the quantum numbers that 
now arise.

\noindent{\it Acknowledgements :}
Many thanks to Steffen Krusch and Nick Manton for useful discussions.
This work was supported by the PPARC special programme
grant ``Classical Lattice Field Theory''.
The parallel computations were performed on COSMOS at the National
Cosmology Supercomputing Centre in Cambridge.
During the preparation of this paper we became aware of independent
work in progress \cite{HoMa} which has a substantial overlap with the
work reported here. There is a qualitative agreement between the two sets 
of results and we thank Conor Houghton and Shane Magee for informing us 
of their work.


\begin{thebibliography}{100}

\bibitem{AN} G.~S. Adkins and C.~R. Nappi, 
The Skyrme model with pion masses,
\textit{Nucl. Phys.} \textbf{B233}, 109 (1984).

\bibitem{ANW} G.~S. Adkins, C.~R. Nappi and E. Witten, 
Static properties of nucleons in the Skyrme model,
\textit{Nucl. Phys.} \textbf{B228}, 552 (1983).

\bibitem{BBT} C. Barnes, W. Baskerville and N. Turok,
Normal modes of the $B=4$ Skyrme soliton,
\textit{Phys. Rev. Lett.} \textbf{79}, 367 (1997);
Normal mode spectrum of the deuteron in the Skyrme model,
 \textit{Phys. Lett.} \textbf{B411}, 180 (1997).

\bibitem{BKS} R.~A. Battye, S. Krusch and P.~M. Sutcliffe,
 Spinning Skyrmions and the Skyrme parameters, 
\textit{Phys. Lett. } \textbf{626}, 120 (2005).

\bibitem{BMS} R.~A. Battye, N.~S. Manton  and P.~M. Sutcliffe,
 Skyrmions and Nuclei, {\em in preparation}.

\bibitem{BS3} R.~A. Battye and P.~M. Sutcliffe, Symmetric Skyrmions, 
\textit{Phys. Rev. Lett.} \textbf{79}, 363 (1997);
Solitonic fullerene structures in light atomic nuclei,
\textit{Phys. Rev. Lett.} \textbf{86}, 3989 (2001);
Skyrmions, fullerenes and rational maps,
\textit{Rev. Math. Phys.} \textbf{14}, 29 (2002). 

\bibitem{BS10} R.~A. Battye and P.~M. Sutcliffe,
Skyrmions and the pion mass,
\textit{Nucl. Phys.} \textbf{B705}, 384 (2005).

\bibitem{BTC} E. Braaten, S. Townsend and L. Carson,
Novel structure of static multisoliton solutions in the Skyrme model,
\textit{Phys. Lett.} \textbf{B235}, 147 (1990).

\bibitem{HoMa1} C.~J. Houghton and S. Magee, 
A zero-mode quantization of the Skyrmion,
\textit{Phys. Lett.} \textbf{B632}, 593 (2006).

\bibitem{HoMa} C.~J. Houghton and S. Magee, {\em in preparation}.

\bibitem{HMS} C.~J. Houghton, N.~S. Manton and P.~M. Sutcliffe,
Rational maps, monopoles and Skyrmions,
\textit{Nucl. Phys.} \textbf{B510}, 507 (1998).

\bibitem{Kop} V. Kopeliovich,
Characteristic predictions of topological soliton models,
\textit{J. Exp. Theor. Phys.} \textbf{93}, 435 (2001).

\bibitem{Kr2} S. Krusch, 
Homotopy of rational maps and the quantization of Skyrmions,
\textit{Ann. Phys.} \textbf{304}, 103 (2003).

\bibitem{book} N.~S. Manton and P.~M. Sutcliffe,
{\em Topological Solitons}, Cambridge University Press (2004).

\bibitem{Sk} T.~H.~R. Skyrme, 
A nonlinear field theory,
\textit{Proc. R. Soc. Lond.} \textbf{A260}, 127 (1961).

\bibitem{Wi} E. Witten, 
Global aspects of current algebra,
\textit{Nucl. Phys.} \textbf{B223}, 422 (1983);
Current algebra, baryons, and quark confinement,
{\it ibid} \textbf{B223}, 433 (1983).

\end{thebibliography}
\end{document}